\documentclass[final,5p,times,twocolumn]{elsarticle}

 \usepackage{graphics}

\usepackage{amssymb}

 \usepackage{lineno}

\journal{NIM A  RICAP-2013}

\begin{document}

\begin{frontmatter}



\title{ Recent VERITAS Results }


\author{D. Staszak, for the VERITAS Collaboration}

\address{McGill University, 3600 rue University, Montreal, Quebec, H3A2T8, Canada}

\begin{abstract}

VERITAS is an array of four imaging atmospheric Cherenkov telescopes near Tucson, Arizona and is one of the world's most sensitive detectors of very high energy (VHE: $>$100 GeV) gamma rays.  
The scientific reach of VERITAS covers the study of both extragalactic and Galactic objects as well as the search for astrophysical dark matter.
In these proceedings we will discuss the status of VERITAS operations and upgrades and present a selection of recent results.

\end{abstract}

\begin{keyword}


\end{keyword}

\end{frontmatter}


\section{Introduction and Array Status}
\label{introSec}

VERITAS is an array of four imaging atmospheric Cherenkov telescopes near Tucson, Arizona and is one of the world's most sensitive detectors of very high energy (VHE: $>$100 GeV) gamma rays. 
The array has been fully operational since 2007 and now has a catalog of more than 40 detected extragalactic and Galactic sources.
VERITAS science covers a wide range of topics. 
In these proceedings we will highlight a few recent results from both extragalactic and Galactic areas of study as well as briefly 
describe the status of the dark matter science program.

\begin{figure}
\begin{center}
\includegraphics[height=45mm,width=50mm]{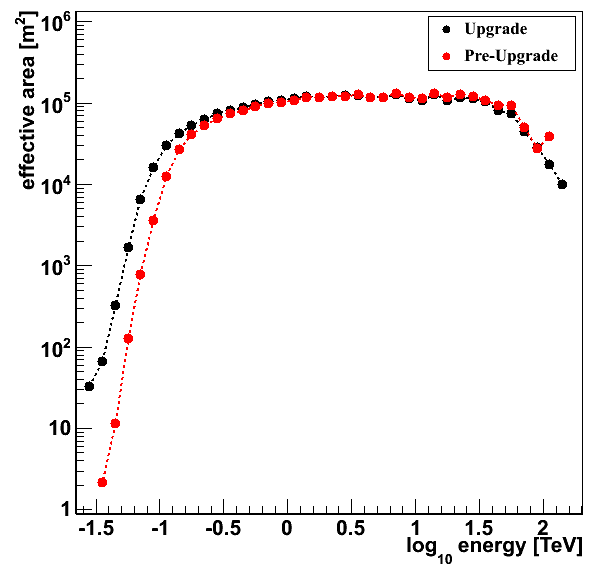}
\caption{Simulated VERITAS effective detection area as a function of primary energy.  Shown is the response with old and new PMTs for $20^{\circ}$ zenith gamma rays. }
\label{upgradePlot}
\end{center}
\end{figure}

The VERITAS array can currently measure astrophysical gamma rays over the energy range of $\sim$85 GeV to 30 TeV with an energy resolution of $\sim$15-25$\%$, an angular 
resolution of $<$0.1$^{\circ}$ at 1 TeV, and a pointing accuracy error $<$50$^{''}$. 
VERITAS can detect a 1$\%$ Crab flux source in $\sim$25 hours and the Crab Nebula itself in $\sim$70 seconds (the Crab Nebula flux is taken as 2.1 $\times$ 10$^{-10}$ $\gamma$s cm$^{-2}$s$^{-1}$).
Two major upgrade efforts have been performed since 2007.
The first in 2009 developed an improved optical alignment tool\cite{mccann} and 
relocated one of the telescopes to better symmetrize the array\cite{perkinsT1}.
The second in 2011-2012 replaced the L2 trigger system\cite{L2} and installed new PMTs in each of the four cameras\cite{PMTs}.
The new PMTs are high quantum efficiency models (with photon detection efficiencies reaching $35\%$) and were installed during the summer monsoon shutdown,
resulting in no observing downtime.
These PMTs collect significantly more light and extend our effective detection area, especially at lower energies.
Fig. \ref{upgradePlot} shows our sensitivity before and after this upgrade\cite{kiedaUpgrade}.

\begin{figure}
\begin{center}
\includegraphics[width=65mm]{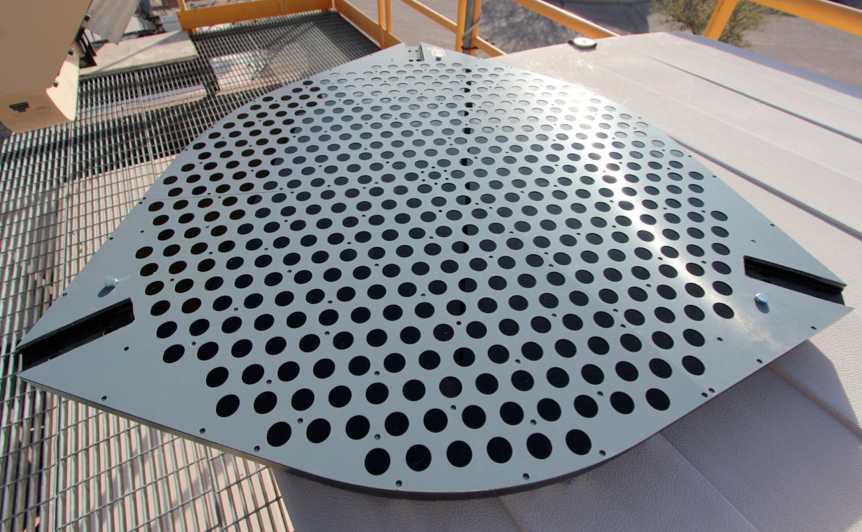}
\caption{One of the filters that is placed over the camera to extend observation into bright moonlight conditions.}
\label{filterPlot}
\end{center}
\end{figure}

Since the latter part of the 2011-2012 observing season, VERITAS has aggressively extended observations to cover a larger 
fraction of the monthly moon cycle.
During bright moonlight phases we now observe in two separate non-standard modes designed to limit the effect of background light introduced by the moon: reducing the PMT high-voltages,
and covering each of the four PMT cameras with UV-filters.
These operation modes increased our effective live time over the season by $\sim$20$\%$.
However, since these observations require dedicated simulations and analysis care, the main use of this time so far is to monitor known or potential sources 
to catch interesting flaring events. 
Fig. \ref{filterPlot} shows one of these filters that is designed to pass the peak of the Cherenkov spectrum ($\sim$250-400 nm) while 
blocking the majority of the moon's reflected solar spectrum.

\section {Extragalactic Results}
\label{extraSec}

All VERITAS extragalactic source detections are blazars with two exceptions: M82, a starburst galaxy, and M87, a radio 
galaxy\footnote{At the 2013 ICRC, VERITAS announced the detection of a second radio galaxy, NGC 1275.}.
The VERITAS extragalactic science program emphasizes the detection and characterization of as many blazars and AGN as possible  since there are many scientific topics that benefit from a sizeable population to study.
VERITAS currently has exposures on roughly 100 different AGN and has a monitoring program to catch interesting events (flares) from a large
number of VHE detected or candidate AGN.

VERITAS blazar source detections range from the relatively close, Mrk 421 with a redshift of 0.030, to the most distant blazar yet detected, PKS 1424+240 
with a newly measured lower limit redshift of 0.6035.
This newly-published limit\cite{pks} makes PKS 1424+240 one of the most intriguing VHE blazar detections thus far and challenges the existing models
of the extragalactic background light (EBL).
A combined spectral analysis of {\it Fermi}-LAT and VERITAS data shows a clear spectral break at $\sim$100 GeV\cite{pks}.
Whereas VHE photons are believed to be absorbed by $\gamma_{VHE} + \gamma_{EBL} \rightarrow e^{+} + e^{-}$ processes, GeV photons are expected
to be unaffected by these interactions.
A deabsorbed VHE spectrum of PKS 1424+240 using several existing EBL models does not adequately correct the majority of the VHE points to an extrapolated LAT spectrum.
While the remaining difference could arise from unaccounted-for effects at the VHE source, it could also indicate the possibility that the gamma-ray opacity of the
Universe has been overestimated.

\begin{figure}
\includegraphics[width=80mm]{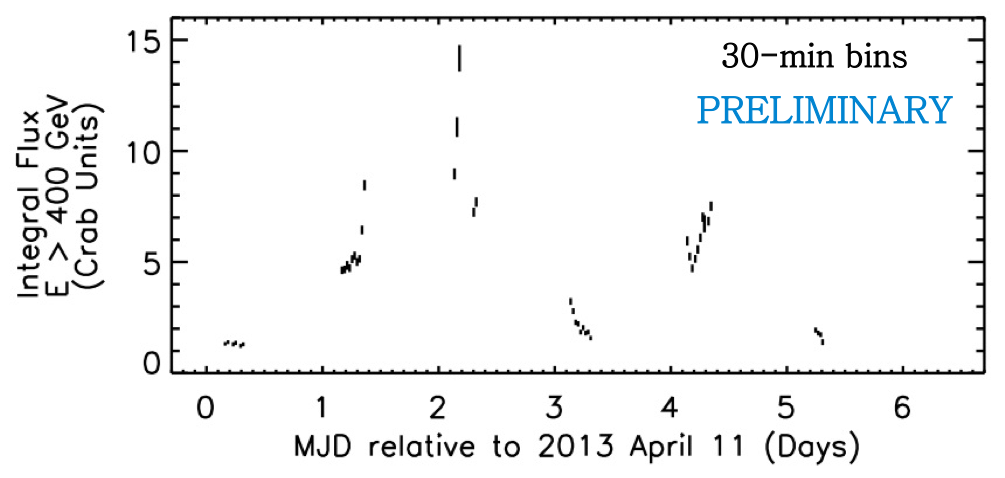}
\caption{Preliminary Mrk 421 integrated flux ($>$400 GeV, 30-minute bins) measured during the historic April 2013 flaring episode.  Fluxes are reported in 
Crab flux units and dates are reported in days relative to April 11, 2013.} 
\label{mrk421Plot}
\end{figure}

This past observing season was witness to a historic flaring event from Mrk 421.
In March of 2013 a MWL campaign began to look at the nearby blazar with MAGIC, VERITAS, and the newly commissioned {\it NuSTAR} satellite. 
Flaring activity was seen by all three, as well as  with {\it Fermi}-LAT and {\it Swift}\cite{atel}. 
Fig. \ref{mrk421Plot} shows VERITAS preliminary flux levels that reach up to $\sim$14 Crab units.
Note that many of the holes in VERITAS coverage are filled by MAGIC (not shown here).
This is a rich dataset containing up to ~11 hours of VHE/GeV/X-ray overlap data (X-rays from both {\it NuSTAR} and {\it Swift}) that has only begun to be explored.  

In addition to the excitement from the two blazars described above, we can report two new VERITAS blazar detections, 1ES 1011+496 and 
1ES 0647+250\cite{dumm}.
Both of these blazers were originally discovered in the VHE band by MAGIC but not detected by VERITAS prior to this season.
Both were also initially observed during partial moonlight observations, emphasizing the usefulness of extending observation into moonlight conditions.
1ES 1011+496 was detected in 10.4 hours at 8.5$\sigma$ significance with an observed flux of 6.3$\%$ Crab units $>$150 GeV. 
1ES 0647+250 was detected in 10.9 hours at 6.2$\sigma$ significance with an observed flux of 7$\%$ Crab units $>$200 GeV. 

\section {Galactic Results}
\label{galSec}

VERITAS is a northern hemisphere observatory and so primarily views the outer galaxy.
Galactic sources detected by VERITAS include SNRs, PWN, binary systems, unidentified sources, and the
first detected pulsar above 100 GeV.

One of the main focuses of the VERITAS Galactic science program is to understand the origin of TeV cosmic rays in the Galaxy.
Gamma rays are unique probes of energized regions since they are neutral and thus not deflected by the intervening magnetic fields.
SNRs have long been believed to be the main source up to energies around the knee.
However, the gamma-ray production mechanism isn't necessarily understood and can vary from source to source (from shock acceleration at the shell to pulsar emission to PWN emission).

\begin{figure}
\begin{center}
\includegraphics[width=59mm]{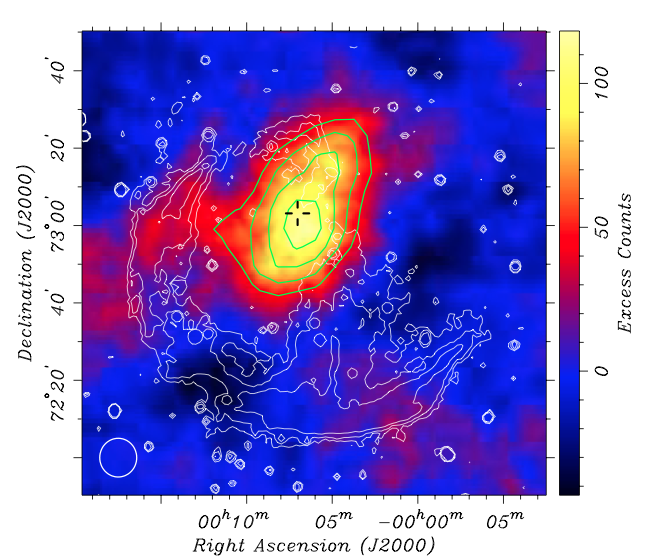}
\caption{Excess counts in the vicinity of CTA 1. The color represents the number of excess counts and the white contours represent 
the radio emission of the shell.  The cross represents the position of the $\gamma$-ray pulsar position measured by $Fermi$-LAT and 
the green contours are the 3$\sigma$, 4$\sigma$, 5$\sigma$, and 6$\sigma$ regions of the VERITAS detection. The circle in the lower 
left shows the VERITAS PSF.}
\label{ctaPlot}
\end{center}
\end{figure}

CTA 1 is a composite shell-type SNR that is X-ray filled and has a radio shell of diameter 1.8$^{\circ}$.
$Fermi$-LAT performed a blind search for pulsations and discovered a $\gamma$-ray pulsar in the center region with a period of 315 $ms$
and an age comparable to that of the SNR\cite{fermiCTA1}.
This was the first direct evidence that this SNR could possibly be a PWN.
VERITAS observed CTA 1 for 41 hours and detected extended emission (with angular extent $\sim$0.25$^{\circ}$) from the object at 6.3$\sigma$ post-trials significance\cite{cta1} (see Fig. \ref{ctaPlot}).
The measured flux is $\sim$4$\%$ Crab flux units $>$1 TeV and the fitted centroid of the VERITAS significance is within 5 arcmin of the $Fermi$ pulsar.
The source name of this object is VER J0006+729.
A PWN explanation is strongly favored by the physical proximity of the VHE source to the $Fermi$ pulsar as well as the extended yet compact emission region.
CTA 1 fits well with the emerging picture of relatively young, high E-dot pulsars being good candidates for TeV PWN emission\cite{karg}.

There are four gamma-ray binary systems detected at VHE energies: LSI +61$^{\circ}$ 303, HESS J0632+057, PSR B1259-63, and LS5039.
These are complicated systems where the VHE emission may arise from either colliding winds or be powered by accretion.   
VERITAS has detected two of the four, LSI +61$^{\circ}$ 303\cite{andyLSI} and HESS J0632+057\cite{hessJ}, the latter being unique of the four in that it isn't detected by the $Fermi$-LAT.

LSI +61$^{\circ}$ 303 is a high mass X-ray binary system that contains a compact object (neutron star or black hole) orbiting a large main sequence star with a 26.5 day elliptical orbit.
It exhibits X-ray emission throughout the orbit, radio emission that peaks at periastron and apastron, MeV/GeV emission throughout the orbit, 
and VHE activity typically detected at apastron (seen by MAGIC\cite{magicLSI} and VERITAS).
The history of VHE emission of this system is intriguing.
Strong detections by MAGIC and VERITAS in 2005-2007 were followed by several years of marginal or non-detections.
In 2010 VERITAS detected the system but at a reduced flux level and near periastron as opposed to apastron.
VERITAS can now report a 11.9$\sigma$ detection in the apastron phase from 25 hours of data taken in 2011-2012, with the object exhibiting a $\sim$10$\%$ Crab units flux ($>$350 GeV)\cite{andyLSI}\cite{newLSI}. 
While it is possible that VERITAS missed the emission from this object during the intermediate years, it is also possible that it undergoes multi-year
variations.

\begin{figure}
\begin{center}
\includegraphics[width=70mm]{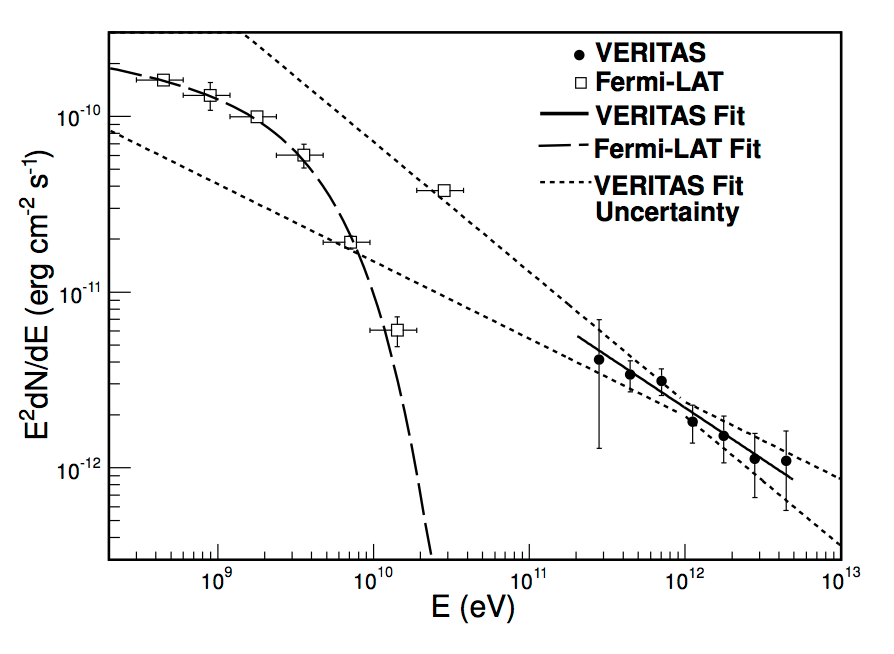}
\caption{Energy spectrum of LSI +61$^{\circ}$ 303 using contemporaneous $Fermi$-LAT and VERITAS data from 2011-2012.
Figure taken from \cite{andyLSI}.}
\label{lsiPlot}
\end{center}
\end{figure}

Fig \ref{lsiPlot} shows a combined $Fermi$-LAT and VERITAS energy spectrum from December 2011 to February 2012. 
$Fermi$ data used in this analysis covered this larger period of VERITAS data collection and was not simply restricted to VERITAS live time.
While the VERITAS data is well fit with a power-law, the $Fermi$ data exhibit a clear spectral break around 4 GeV.
This cutoff behaviour is suggestive of the larger population of GeV pulsars discovered by $Fermi$ and may indicate different emission populations in GeV
and TeV.
However, note that the $Fermi$ data isn't strictly simultaneous, so it is possible there is short-term variability in GeV that we are simply not sensitive to.

\section {Dark Matter Results and Outlook}
\label{dmSec}

The search for particle dark matter is a major effort within VERITAS.
Many of the leading candidates for dark matter, including WIMPs, axions, and Kaluza-Klein particles,
predict annihilation and/or decay channels with photon final states.
These photons can arise from either direct annihilation (for example, $\chi \chi$$\rightarrow$$\gamma \gamma$, in the case of the supersymmetric neutralino)
or from hadronic or leptonic decay chains (leptonic channels with final state radiation).
Direct annihilation would provide the cleanest evidence of DM but is suppressed relative to hadronic channels by higher order loops.
VERITAS and other IACTs are an important part of the search for DM since they are sensitive to photon energies that are relatively unconstrained by
direct nuclear recoil experiments and collider experiments.
Further, any hint of a DM particle seen on Earth would need to be confirmed as the actual astrophysical DM.

VERITAS has targeted several different DM over-densities, including galaxy clusters, dwarf spheroidal galaxies (DSphs), the Galactic Center (GC), 
and candidate unidentified {\it Fermi}-LAT sources.
Each source class has advantages and disadvantages based on predicted $\gamma$-ray flux, distance from Earth, astrophysical background levels,
and DM density.
For brevity in these proceedings, we'll discuss results and projections from two of the most promising 
observations, the GC and DSphs.

\begin{figure}
\begin{center}
\includegraphics[width=70mm]{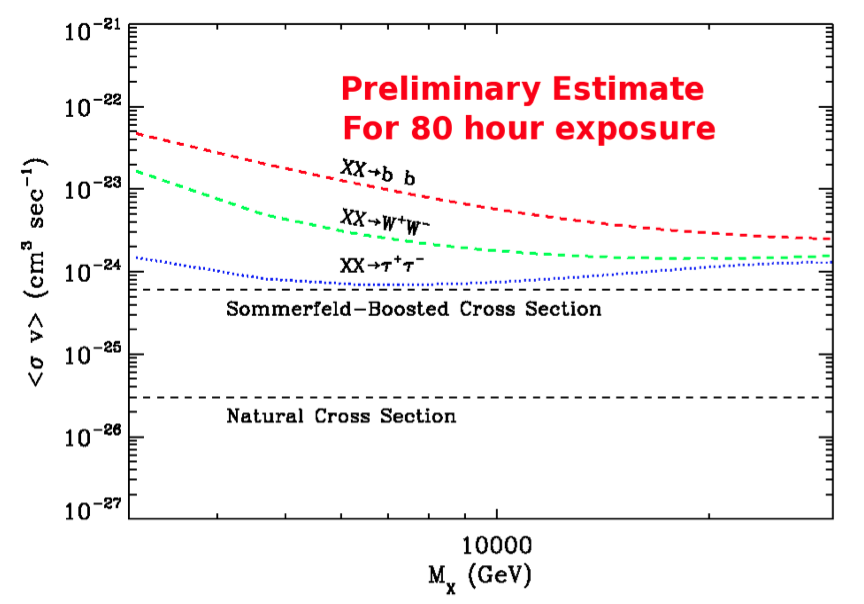}
\caption{Projected VERITAS DM sensitivity limits of the GC using data through the end of the 2013 observing season.
Figure taken from \cite{andyDM}.}
\label{galCenPlot}
\end{center}
\end{figure}

The GC is a challenging region to analyze because of the dominant astrophysical
backgrounds.  
Additionally, the GC is only visible at large zenith angles (LZA: zenith$>$$50^{\circ}$) since VERITAS is a northern hemisphere observatory. 
VERITAS uses an ON and OFF observation technique to better characterize the background, whereby 
observations are split between directly targeting the GC and targeting a field in the vicinity of GC without a known VHE emitter.
VERITAS detects the astrophysical object at the GC (Sgr A*)\cite{GCpaper} and finds a spectrum that is in agreement with 
prior measurements of this source from other experiments\cite{whippleGC}\cite{hessGC}.
The DM search strategy in this region is motivated by line-of-sight integrals of the DM density that suggest overdensities extending
off the Galactic Plane (GP).
Signal and background regions are selected within the field of view that avoid the known emitters and the GP while optimizing to
higher DM density areas (see \cite{andyDM} for further details).
Fig. \ref{galCenPlot} shows DM sensitivity projections with VERITAS data including the current 2012-2013 observing 
season\cite{andyDM}.
Note that LZA observations of this region increase the array collection area at high energies.
This could benefit VERITAS since the LHC has seen no new physics in their 7-8 TeV Run and the relatively high mass Higgs they measure
may have the effect of increasing the SUSY production energy scale, leading to higher mass neutralinos.

\begin{figure}
\begin{center}
\includegraphics[width=67mm]{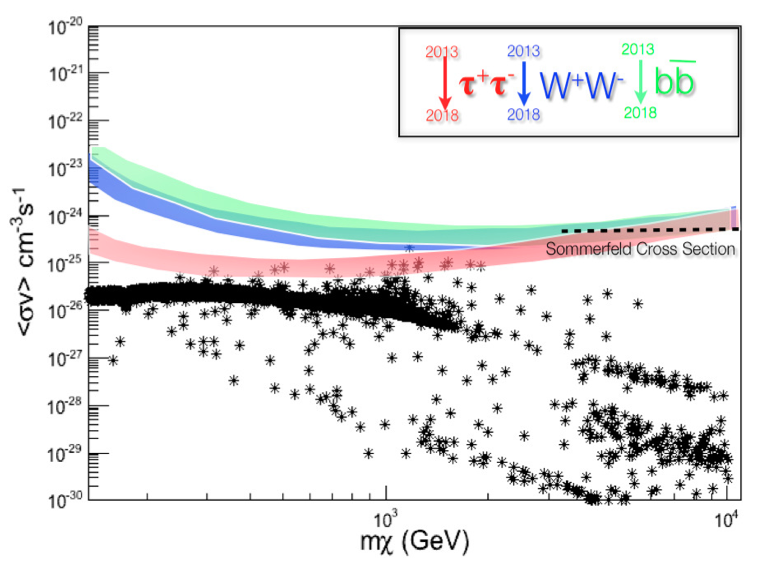}
\caption{Projected VERITAS DM sensitivity limits from a stacked DSph analysis using data collected through the end
of the 2013 observing season (top of each band) and data expected through the 2018 observing season (bottom of each band).  
Black points represent instances of a typical MSSM model and the dashed line represents the predicted level of the DM cross section boosts
from Sommerfeld-type scenarios\cite{sommerfeld1}\cite{sommerfeld2}.  Figure taken from \cite{andyDM}.}
\label{dwarfPlot}
\end{center}
\end{figure}

DSphs are very promising observation targets because they are DM dominated objects that are nearby with 
little astrophysical background.
VERITAS has observed several DSphs and previously published upper limits on DM flux from Draco, Ursa Minor, Willman 1, and Bo\"{o}tes 1\cite{dwarfs}
and more recently from a deeper exposure of Segue 1\cite{segue1}.
Current limits from these studies are two orders of magnitude away from constraining the canonical models of DM.
However, Segue I limits do constrain the available phase space of some boosted annihilation scenarios (e.g. Sommerfeld mechanism or leptophilic models).
In Fig. \ref{dwarfPlot} we show predicted sensitivity improvements based on additional data collected through the end of 2013
and the use of an improved analysis technique from \cite{stacking}.
Projections are also shown for the expected data collected through 2018.
This new technique combines the data from all DSph observations and weights each event based on the measured DM density function.
It has been shown to improve limits on DSphs measured by {\it Fermi}-LAT and will be the first stacking analysis of IACT 
data.

\section{Conclusion}

VERITAS is yielding a wide range of results covering extragalactic and Galactic science topics as well as the search for particle DM.
An improved DM analysis is underway, and DM is a high priority for the coming years.
VERITAS reports the detection of two new blazars, the first new detections with the upgraded camera.
Fortuitous timing of a MWL campaign led to the detection of a historic flare from Mrk421.
Preliminary VERITAS flux levels of this flare reached 14 Crab units and a good fraction of the data has X-ray, GeV, and TeV overlapping coverage.
VERITAS reports the detection of extended emission from the central region of CTA 1 and finds the emission likely to arise from a PWN.
Utilizing 2011-2012 MWL data we show a contemporaneous GeV-TeV spectrum of the high mass X-ray binary object LSI +61$^{\circ}$ 303.
Results show a spectral break at a few GeV and seem to indicate two populations of emitters.  
However, further investigation will be needed to form a definitive understanding of this object's nature.
Finally, we report that the recent series of upgrades resulted in no loss of observing time and have successfully lowered the energy threshold of the VERITAS array.
Additionally, new bright moonlight observation modes are increasing the amount of observing data per season by $\sim$20$\%$.

As a final note we can say that for the first time an IACT experiment has opened up a fraction of its observing time to the larger community.
A VERITAS/Fermi pilot program was initiated for Cycle-6 of the Fermi Guest Investigator program
(2012-2013)\footnote{For more information see http://fermi.gsfc.nasa.gov/ssc/proposals/}.
Roughly 4$\%$ of the accepted GI proposals were joint Fermi/VERITAS proposals. 
VERITAS observations for these proposals will take place during the next observing season.  

\vspace{-0.2in}
\section*{Acknowledgment}
\vspace{-0.1in}
This research is supported by grants from the U.S. Department of Energy Office of Science, 
the U.S. National Science Foundation and the Smithsonian Institution, by NSERC in Canada, by 
Science Foundation Ireland (SFI 10/RFP/AST2748) and by STFC in the U.K. We acknowledge the excellent
work of the technical support staff at the Fred Lawrence Whipple Observatory and at the collaborating 
institutions in the construction and operation of the instrument.
\vspace{-0.1in}





\bibliographystyle{elsarticle-num}
\bibliography{<your-bib-database>}



\end{document}